# Attention is All You Want: Machinic Gaze and the Anthropocene

Liam Magee[1]
Vanicka Arora[2]

[1]Western Sydney University, Australia
l.magee@westernsydney.edu.au
[2]University of Stirling, United Kingdom
vanicka.arora@stir.ac.uk

May 2024

## Abstract

This chapter experiments with ways computational vision interprets and synthesises representations of the Anthropocene. Text-to-image systems such as MidJourney and StableDiffusion, trained on large data sets of harvested images and captions, yield often striking compositions that serve, alternately, as banal reproduction, alien imaginary and refracted commentary on the preoccupations of Internet visual culture. While the effects of AI on visual culture may themselves be transformative or catastrophic, we are more interested here in how it has been trained to imagine shared human, technical and ecological futures. Through a series of textual prompts that marry elements of the Anthropocenic and Australian environmental vernacular, we examine how this emergent machinic gaze both looks out, through its compositions of futuristic landscapes, and looks back, towards an observing and observed human subject. In its varied assistive, surveillant and generative roles, computational vision not only mirrors human desire but articulates oblique demands of its own.

*Keywords* – Visual analysis, visual cultures, media studies, generative AI, psychoanalysis

# Latent Constellations of the Anthropocene

In the disappointments that followed failed predictions about the imminent arrival of artificial intelligence in the 1950s, one consequence was a sustained concentration on the more tractable problems of computer vision. At the University of Utah, where the Defence Advanced Research Projects Agency (DARPA) funded a decade of research from 1965 until 1975, researchers produced vital foundations for the computer graphics industry, and several would themselves later launch commercial start-ups like Adobe and Pixar. Examining the simulation of light, shade, reflection and movement, this early period looked to translate Renaissance innovations in perspective and chiaroscuro into the new medium of the screen (Appel 1968). Though hardware lagged long behind, scanned black-and-white stills of illuminated geometric primitives from the doctoral theses of Gouraud (1971) and Phong (1975) anticipate the developments of computer-generated imagery, computer-aided design and 3D gaming. Responding to demands for faster and more realistic rendering, by the 1990s companies like NVidia had begun to design and market dedicated computer graphics cards to a nascent gamer and professional graphics market.

In a strange kind of return, it was the affordances of these cards that much later, by the late 2000s, led to efficiencies in the field of machine learning, or what today has become, in a different semantic kind of renaissance, known again as Artificial Intelligence. Graphics cards excel in the parallel execution of simple repetitive mathematical instructions, like the transformation of still frames of pixels to simulate a moving camera, and this same facility can be used to distribute the iterative processing of large training sets, to learn approximations of the structure of text, image and other media. With the release of text-to-image models such as Midjourney, DALL-E, and Stable Diffusion and large language models such as ChatGPT, 2022 witnessed a watershed moment in the sophistication and fidelity with which these approximations could generate apparently novel outputs, progressing from research curiosities and social media showpieces to instruments that intrude upon wider circuits of cultural production and consumption.

One line of questioning posed by the emergence of these systems and their accompanying impact on circulating scopic regimes and visual cultures – alongside issues of property theft, plagiarism and bias – relates to the shift they effect in conceptualising visual culture. Critical engagements have focused on the unoriginality and plagiaristic qualities of generative AI. Steyerl for instance has called its outputs 'mean images' (2023), punning on its tendency to average out and to reproduce, mean-spiritedly, the work of human artists. Chiang termed ChatGPT as 'an Internet with blurry edges' (2023), while Salvaggio (2024) has recently elaborated upon the de-noising and diffusional operations of text-to-image models. Across each of these critiques, a technical language of approximation (mean, compression, noise, diffusion) is brought back analytically to demarcate a set of *cultural* operations performed by these machines. What is imagined is a kind of vast, sequenced and extractive compression of the work of human artists, photographers, social media users and machine learning engineers who, at different times and of course with different impulses, create, design, circulate and synthesise visual media into generative algorithms.

Without relinquishing this critical charge, we also want to argue that this synthetic reflection is not a simple mirroring of a collective and diffused human apprehension of the visual. More is involved, in other words, than the mechanical reproduction of a collective, synthesised perception, or perhaps to put it another way, this is a reproduction that to an extreme 'can capture images which escape natural vision' (Benjamin, 1986). As an artificial vision, the technical production of images here involves a different and inhuman gaze, creating associations of form, aesthetics and genre that reconstitute rather than reflect prior visual cultures. Machinic vision, a mode of perception described by Virilio (1995) and Johnston (1999) in context of photography, cinematography, and earlier forms of digital technologies, allowed for the proliferation of disembodied perspectives contained within the nonhuman, rendering visible forms, aesthetics and patterns imperceptible by the human eye. In the context of machine learning, Mackenzie and Munster (2019) have described the general synthetic production of images via algorithms as a shift toward a mode of 'invisuality', a neologism that conveys their inward-looking quality. Rather than responding to a set of physical, externalised, or cultural conditions, image synthesis relies exclusively upon the intricate mappings of language terms and clusters of pixels established through the system's training. Image synthesis constitutes an inhuman kind of 'platform seeing' (Mackenzie and Munster, 2019). For certain algorithms, ironically, this seeing is learned through reference to auditory perception: the controlled addition of noise to source data, which is reversed during application, as a randomised image is de-noised into a recognisable approximate to input text (Salvaggio, 2024).

At times this distortion can certainly be seen as normative, tending towards a mean or central tendency of representation, collapsing the eclecticism of cultural forms into a singular stereotype. At other times, less frequently, it veers towards extremes, accentuating absurdist or surrealist features in what could pass for algorithmic 'interpretation'. If we imagine that the machine learns representations by devising a series of axes upon which to distribute distinctive features of text and images – nouns and verbs, for instance, or cats and dogs – we can think about how a textual input (better known as a 'prompt') acts like a guide or map into the intricacies of how these differences are distributed. Each word or token in the prompt – and often chains of words grouped as clauses – arrives upon a location or vector associated with patterns of pixels which then are retrieved to contribute to the eventual image. One example would be a simple linear or geometric form, such as a horizontal line, while another might be the proper name of an artist – and as we show below, proper names do interesting work, not only transferring stylistic properties but perturbing the selection, location, and periodisation of the final image subject. These vectors together cut across the model's multidimensional latent space to form a kind of imaginary figure, an outline or constellation, each part of which maps to a mediatic fragment that is brought forward or synthesised in the generated output. To be effective, the final result should produce no surprise, or in the language of machine learning, low perplexity, and this characteristic reinforces the sense that these machines are necessarily conservative.

## Anatomy of a Prompt

What does the machine see? What forms of visual culture emerge when generative AI processes, curates, and synthesises media from collective archives? Building on Lacan's conceptualisation (1998[1973]), we have previously described this machinic gaze as something that, like the skull in Holbein's *The Ambassadors*, is *anamorphic* to human perception (Arora, Magee and Munn, 2024). Complementing that study, in this instance we focus on the latest version of Midjourney (6 alpha), a generative image subscription service that attends to the parts of an input prompt with impressive detail. As users of this service since version 3, we had noted how prompting becomes more heuristic, more dependent on practice, and on recognition of how models have internalised representations of visual culture. To reign in scope, we opted for a controlled prompt syntax and set of parameters, referencing the Anthropocene and the Australian imaginary, traversing urban, rural, built, and natural landscapes. To begin with we selected the medium of photography, as much of Midjourney's training data, distilled from Instagram and gallery archives alike, is photographic, and we wanted a banal realism, coupled with common Australian sites, to anchor other effects we would later add.

In keeping with a certain psychoanalytic interest, we also choose mid-century Australian surrealists as part of these effects and qualifying influences. While pre-dating the coining of the 'Anthropocene', artists like Joy Hester, Sidney Nolan and Arthur Boyd responded to a fast-changing relation between the human, technology and environment, and now lionised through decades of commercial and institutional curation, these proper names also act as keys – though, as we discuss below, often poorly fitted ones – to visual associations of the Anthropocene. As further probes, we also incorporate direct references to vision technologies – drones, cameras, eyes, surveillant robots – as contemporary manifestations of a long technological, capitalist and colonial tendency, and as increasingly profound appropriators of rare earth materials, fossil fuels, water, land, and cultural property. Finally, we leaned on examples and advice drawn from the Midjourney Discord users community alongside our own experimentation to add elements – 'double exposure', 'superimposed on', 'anamorphic lens flare', 'movie poster collage' – that modified or filtered the composition, drawing out more esoteric aspects of the archive.

To tether these unwieldy elements and to constrain the time spent in trial and error, we designed a simple 'grammar' or structure that delegates roles to parts of a prompt, such as '[medium] [subject] [style] [aspect ratio]'. We expressed this grammar in an old 1950s specification format designed for early computing languages, the Backus-Naur form (BNF), which we then fed into ChatGPT as part of a simple prompt generator.

This helped to constrain the natural language variation format of the prompt to a kind of technical formalism that could serve as both pattern for experimentation ('vary the style while keeping medium and subject constant') and reminder that the 'naturalness' of the prompt is itself illusory. A simple prompt grammar expressed in this way could be:

    <image_prompt> ::= <medium> "of a" <subject> ", " <style> " --ar 3:2"

    <medium> ::= "photograph"

    <subject> ::= <adjective> <noun> | <noun>

    <adjective> ::= 'vibrant' | 'detailed' | 'abstract' | 'modern'

    <noun> ::= 'landscape' | 'cityscape' | 'person' | <site>

    <site> ::= 'Sydney Harbour Bridge' | 'Great Barrier Reef' | …

    <style> ::= <art_movement> ', ' <artist>

    <art_movement> ::= 'impressionist' | 'minimalist' | 'photorealistic' | 'pop art'

    <artist> :: <artist> ::= 'George Barron Goodman' | 'Frank Hurley' | 'Max Dupain' | 'Tracey Moffatt'

Read from top to bottom, the <image_prompt> comprises multiple clauses, such as <medium>, which in turn can be a 'photograph', or extended to include other options. Each clause can be extended indefinitely as lists; and clauses can be added to produce a more rich and expressive grammar, which conversely constrains and controls Midjourney's outputs. We employ variations on this grammar in the four explorations that follow, first testing a set of minimal variations, then adding visual subjects and layered effects as a series of punctuated verbal clauses, and finally using Midjourney's controls to zoom in and expand a single image in multiple directions, to examine how both image and text condition the system's extrapolations.

## Explorations

## 1. Superimpositions

In the first image set, we used the simple grammar outlined above. Fixing medium, subject and aspect to 'photograph of Circular Quay … --ar 3:2', the inclusion of a comma and a single photographer name (*Figure 1*) cues a simulated – and very approximate – temporal shift from the mid-19th century to the late 20th. The top left image contains the anachronism of the Opera house, and all images contain strange spatial reorientations, as though they are the impressions of a vague, imperfect memory. In these abridged examples, the name of the photographer is less a key retrieving records from an archive of images, and more a master signifier conjuring up associations that together manufacture an aura or 'vibe' – steam mixed with image fade perhaps in the case of Goodman and Hurley, internationalist skyscrapers and sharp lines for Dupain, motion blur and contrast for Moffatt – in the final image composite. The perturbations introduced by names are curious. Compared to just 'photograph of Circular Quay', *any* proper name shifts output towards artifice: long or double exposure, periodisation, soft or sharp focus. The collection of names here not only reproduces an aesthetic but reveals a certain distorted history of place: a colonial port, transformed progressively by coal, iron, oil, steel, and electricity, and captured with evolving photographic technique and aesthetic desire. The Opera House and Harbour Bridge figure regardless, their endurance across periods signalling their ubiquity in the archive.

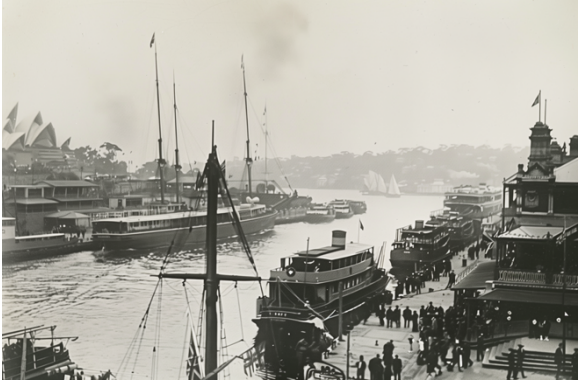

"George Barron Goodman"

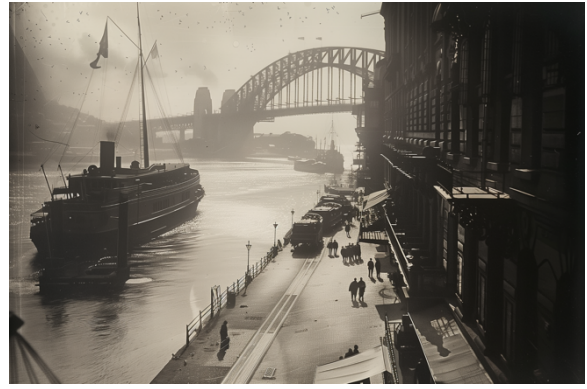

"Frank Hurley"

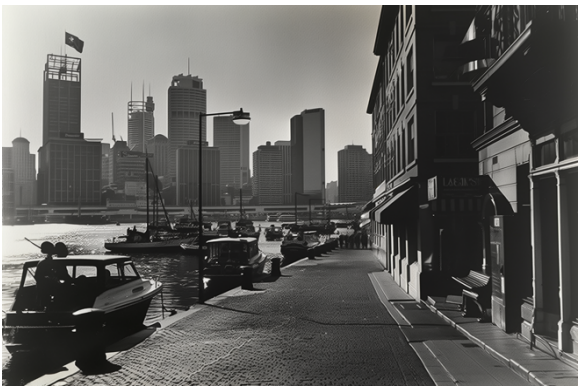

"Max Dupain"

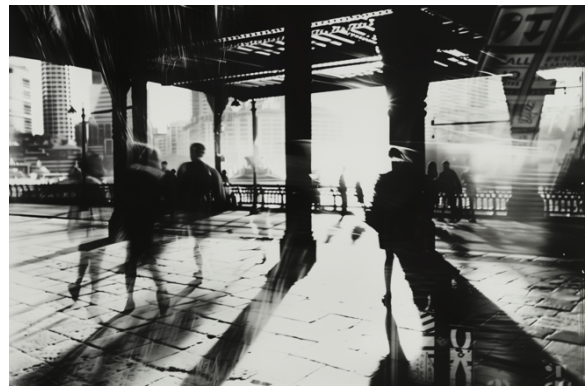

"Tracey Moffatt"

*Figure 1(a-d): "Photograph of Circular Quay"*

## 2. */imagine* the Anthropocene

We expanded this micro-formalism to incorporate a wider 'palette', or set of prompt parts, that fuses together conceptual preoccupations of the Anthropocene, visual culture and photographic technique. The first augmentation is the addition of "double exposure b&w" to "photograph", and two rather than one subjects. This brings into play a split between a background scene, which relates to a familiar part of the Australian landscape, and a superimposed element that reflects some contrastive technological or environmental element. We add adjectives to primary and secondary subjects to strengthen the suggestion of juxtaposition, and extend the stylistic markers to reference time of day, "anamorphic lens flare" and other photographic effects. We also add a nominal and incomplete list of noted Australian surrealist painters ("james gleeson", "joy hester", "arthur boyd", "john perceval", "sidney nolan", "albert tucker"). As well as off-setting the main photographic instruction with additional painterly distortion, this reference to surrealists introduces less their direct style and more, again, of a general ambience of alienation or "out-of-jointness" that in a certain sense stamps the influence of a digitally memorialised artistic movement onto outputs of this new machinic modality.

We submitted this revised specification to ChatGPT to create a miniature engine for producing controlled prompts for Midjourney: one form of media generation feeding into another. We added further instructions: 'use these BNF instructions as a rough guide to produce prompts, given a primary and secondary subject', and then ask ChatGPT to make a elaborate prompt for a supplied topic. For example, we ask:

> Create a prompt about Sorrento back beach overrun by a squad of drones.

And receive in return:

> double exposure b&w photograph of a squad of spectral drones faintly superimposed on a desolate Sorrento back beach, dawn, anamorphic lens flare, arthur boyd, brushstroke --ar 3:2 --s 750

The top left image (*2a*) in the *Figure 2* panel is one of the suggested outputs of this prompt. The other examples in this series of four emphasise similar machinic and scopic themes, retaining certain elements and occasionally introducing variation by hand. We ran versions of these prompts many times over, picking and asking Midjourney to enhance images that initially surprised us with the form of computational interpretation.

Not every term in each prompt makes its way into the conditioning of the image, with levels of suggestion varying depending on order, number of terms or clauses and often, the affinity between clauses. Sometimes an artist's name adds emphasis to the filtering and distortion; at others, it conditions the gender of the android figure embedded in the image. The superimposition can create changes in scale, most evident in the bottom left image, where a human eye is suspended over the shoreline. It can also appear virtually imperceptible, as with the bottom right image, where the pixelated self-portrait is etched into the foreground of a scene of the Australian bush. The bottom right image (*2d*) illustrates the imperfections still evident in rendering reflections, and hints in the prompt subdue the default cinematic and masculine rendering.

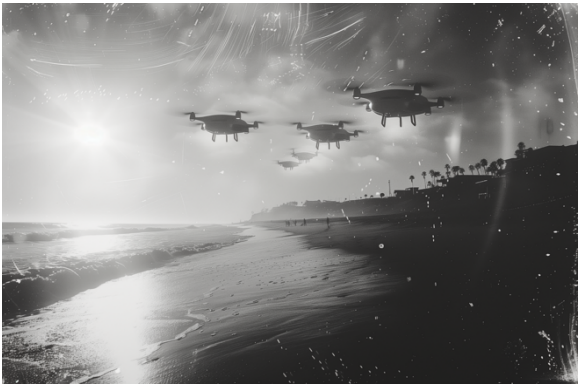

*2a*: "double exposure b&w photograph of a squad of spectral drones faintly superimposed on a desolate Sorrento back beach, dawn, anamorphic lens flare, arthur boyd, brushstroke"

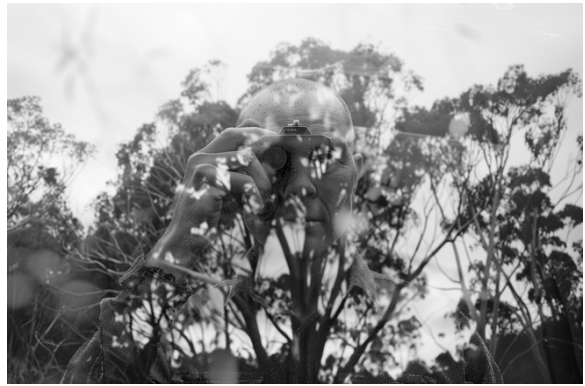

*2b:* "double exposure b&w self-portrait of a spectral humanoid bot, direct frontal pose, taking a photo, close-up, head and shoulders only, faintly superimposed on mirrored reflection of Australian bushland, early morning, brushstrokes, dignified poses"

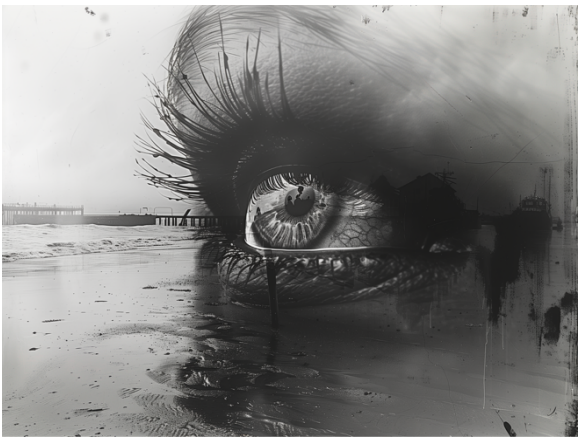

*2c*: "double exposure b&w photograph of enormous translucent eye looking in a mirror reflecting serene apocalyptic 2020s Sorrento back beach, midnight, anamorphic lens flare, James Gleeson, Joy Hester, brushstroke"

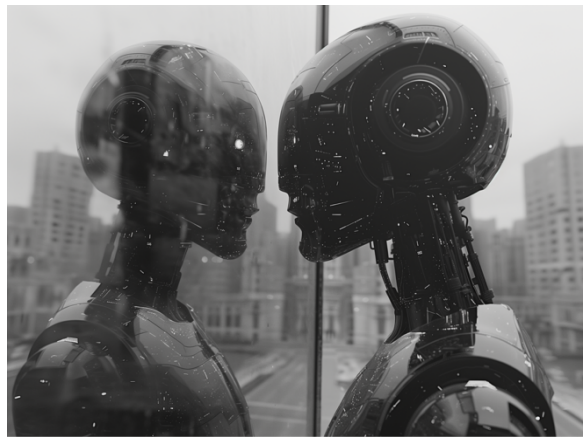

*2d*: "b&w selfie of a robot looking in a mirror reflecting decaying urban background, early afternoon, anamorphic lens flare, joy hester, pixelated, movie poster collage"

*Figure 2(a-d)*

## 3. Multiplying the subject

In the first two explorations, our focus was on prompt grammars and the creation and expansion of a prompt palette, where photographic subject, technique, and style coalesce to generate images that highlighted the tendencies of the machine to fetishise certain aesthetics and styles, and to fixate on specific versions of the subject in the image. Across multiple prompts and responses, the machinic synthesis could cause elision of multiple subjects, or an omission, relating to differences in clause structure. In the third exploration, we stretched some of these tendencies. From initial prompts, we edited out additive clauses, and re-introduced similar subjects and style phrases, creating deliberate additional ambiguities in the prompt. Our first prompt read:

> SUBJECT: underwater b&w photograph of a deep sea diver watching a plastic bottle exploding on decaying dying Great Barrier reef, STYLE: Australiana, Australian Gothic, sydney nolan, joy hester

This led to a similar scene across all four image options: an underwater shot with some form of an explosion vertically distributed across the image, an overhead focused light-source ostensibly beyond the level of the water surface, and a single figure of an undersea diver. The plastic bottle is visible in three of the four outputs, and various forms of coral reef formations are distributed in each of the images. Image 3c (bottom left) of this set of images depicts a plastic bottle *as* the deep-sea diver, illustrating, as in previous examples, the distortions produced between prompt and image.

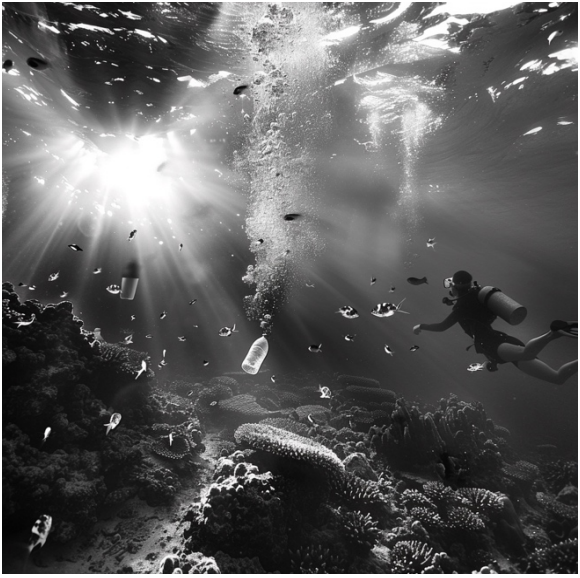
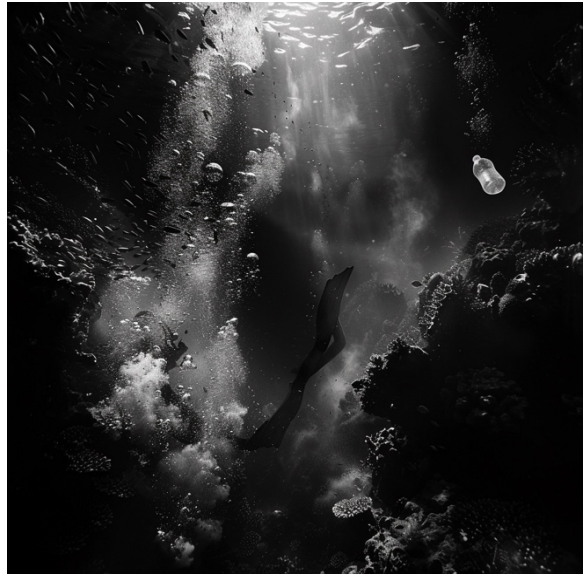
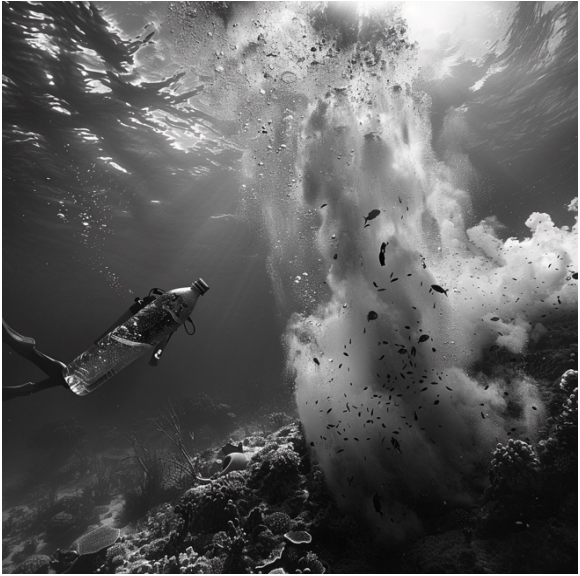
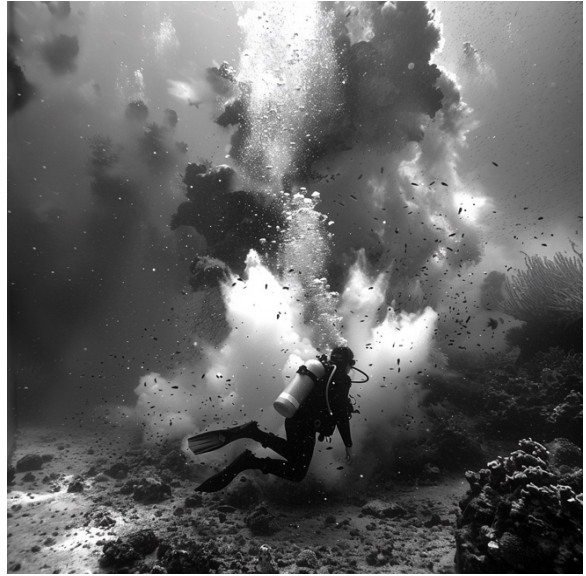

*Figure 3(a-d)*

We then edited as follows:

> SUBJECT: double exposure underwater b&w photograph of futuristic deep sea diver [watching a] **plastic** bottle [exploding on,] **superimposed on** decaying dying Great Barrier reef **made of oversized plastic debris**, STYLE: **Australiana**, **Australian Gothic**, sydney nolan

The elimination of an additive phrase ('watching a' or 'holding a') between the two subjects 'deep sea diver' and 'plastic bottle' meant that new degrees of interpretation were introduced in the image synthesis. The removal of additive phrases resulted in the subversion of structural hierarchies. Renderings of the edited prompt showed, as expected, much greater variation between the four outcomes, though the cinematic quality of the images is retained. In *3e* (top left), the plastic bottle is integrated into the diver's breathing gear, as is the surrounding coral reef and debris. The plastic battle disappears entirely in *3f* (top right), and the debris and coral coalesce. The figure of the diver is supplanted by a single plastic bottle in *3g* (bottom left) as it rests within the coral. Finally, *3h* (bottom right) shows a figure in divergear with a plastic bottle in place of an oxygen tank, standing and observing a coral reef wall enmeshed with debris.

The removal of the additive clauses in the second set of images has the inverse effect of producing greater emphasis on whichever subject is identified by the machine. The diver/bottle is foregrounded in each of the images, with the resultant effect of centering a fictional human subject/object within the context. The verb 'superimposed' scales the image to focus on the diver rather than the reef, with the result that the metaphorical quality of the second set is made more explicit, drawn out by the deliberate slippage in prompt grammar.

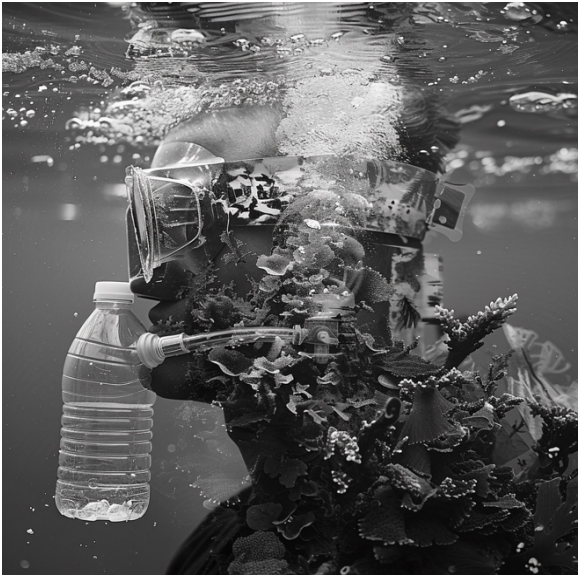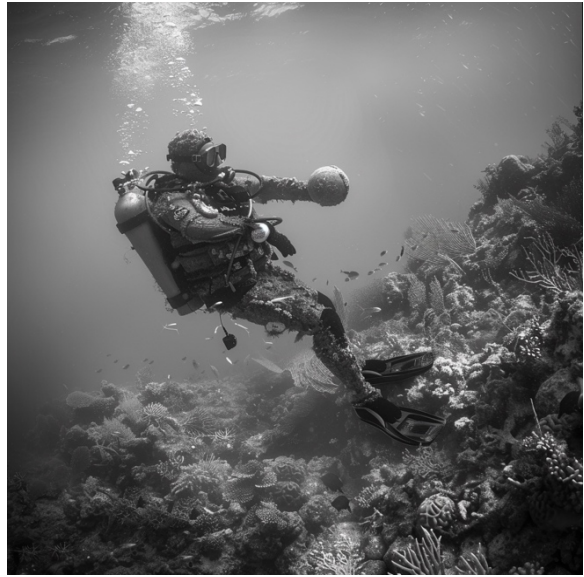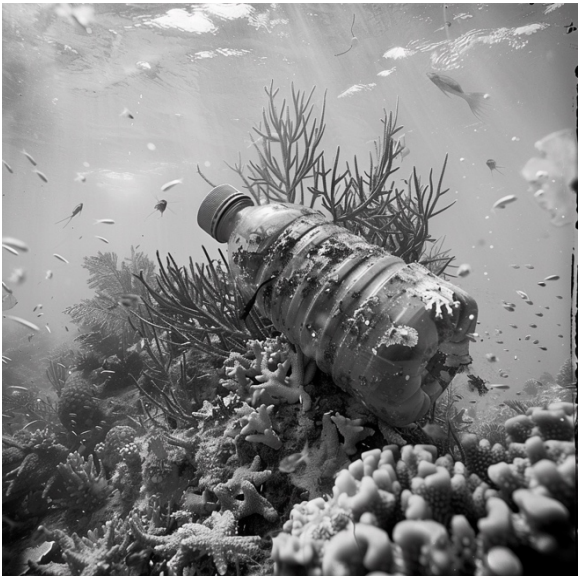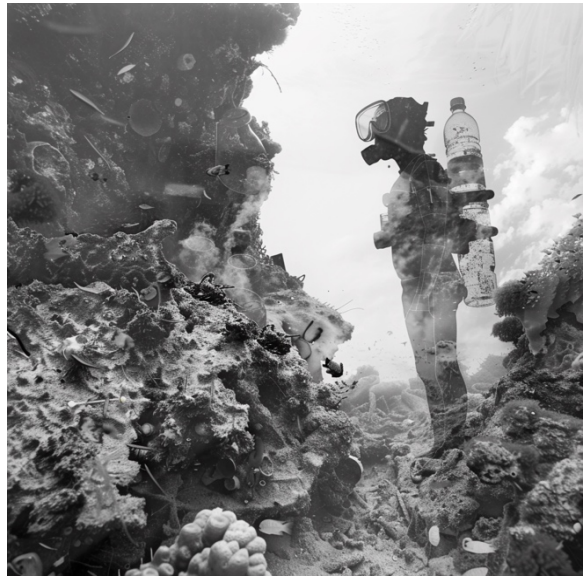

*Figure 3(e-h)*

# 4. Expansions: Extending the Canvas / Mining the Archive

In the fourth set of explorations, we extend images using Midjourney's zoom and pan commands. Both functions work on an existing image by re-submitting the prompt alongside the image and extending the image in multiple directions (**zoom out** 1.5, 2.0, custom zoom) or along a single direction (**pan** left/right/top/bottom). The edges of the image act as suggestive directions for individual elements: a cloud is extended, a landscape forms, and horizon lines are continued across the enlarged frame. New elements get introduced, depending on the existing image and the prompt, but often, in extending the image, the diffusion process creates additional subjects in the image, veering away from the cues supplied in the prompt.

For instance, the inset image in *4a* is one of four images generated images using the prompt:

> SUBJECT: black and white aerial photograph of an Australian farmer watching his farmstead and sheep, a raging fire underground and fireballs in the sky and horizon, symbolising climate change, STYLE: Australiana, Australian Gothic, juxtaposition, Margaret preston

An Australian farmer in the foreground of the image watches his farm burn violently in the distance, with three clouds of smoke originating from the ground and pieces of debris in the air. Upon re-prompting this image through multiple pan commands (in two directions: top, right), an enlarged image is produced, responding to both the preceding image and the original prompt. A second figure now appears on the right, another farmer, positioned as if to watch the original figure, while the debris from the original image is multiplied, enlarged and foregrounded, turning into clearly depicted fireballs and airborne sheep on the newer portions of the virtual canvas. The horizon line and lighting directions are maintained in the new image, but the emphasis and composition has shifted. The pixelation of what was ostensibly a fire cloud in the background of the first image is clarified into a different subject: surreally suspended sheep. The fireballs, absent in the original image, are introduced in the expansion, as the prompt directs the placement of specific subjects in the frame. The image itself shifts genre: from a poignant portrait of a farmer witnessing climate change to a fantastical dystopia, complete with a second observer figure.

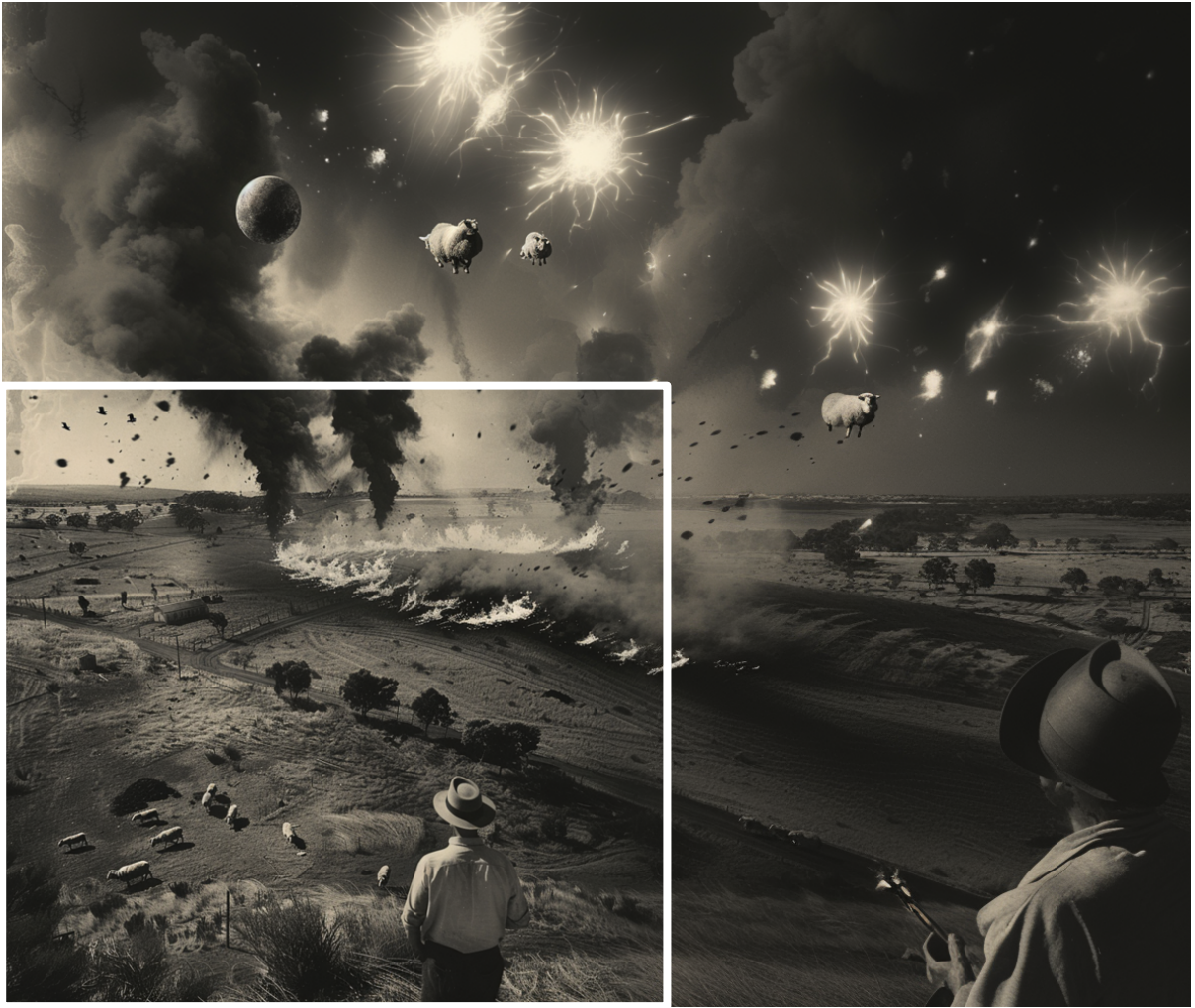

*Figure 4a*

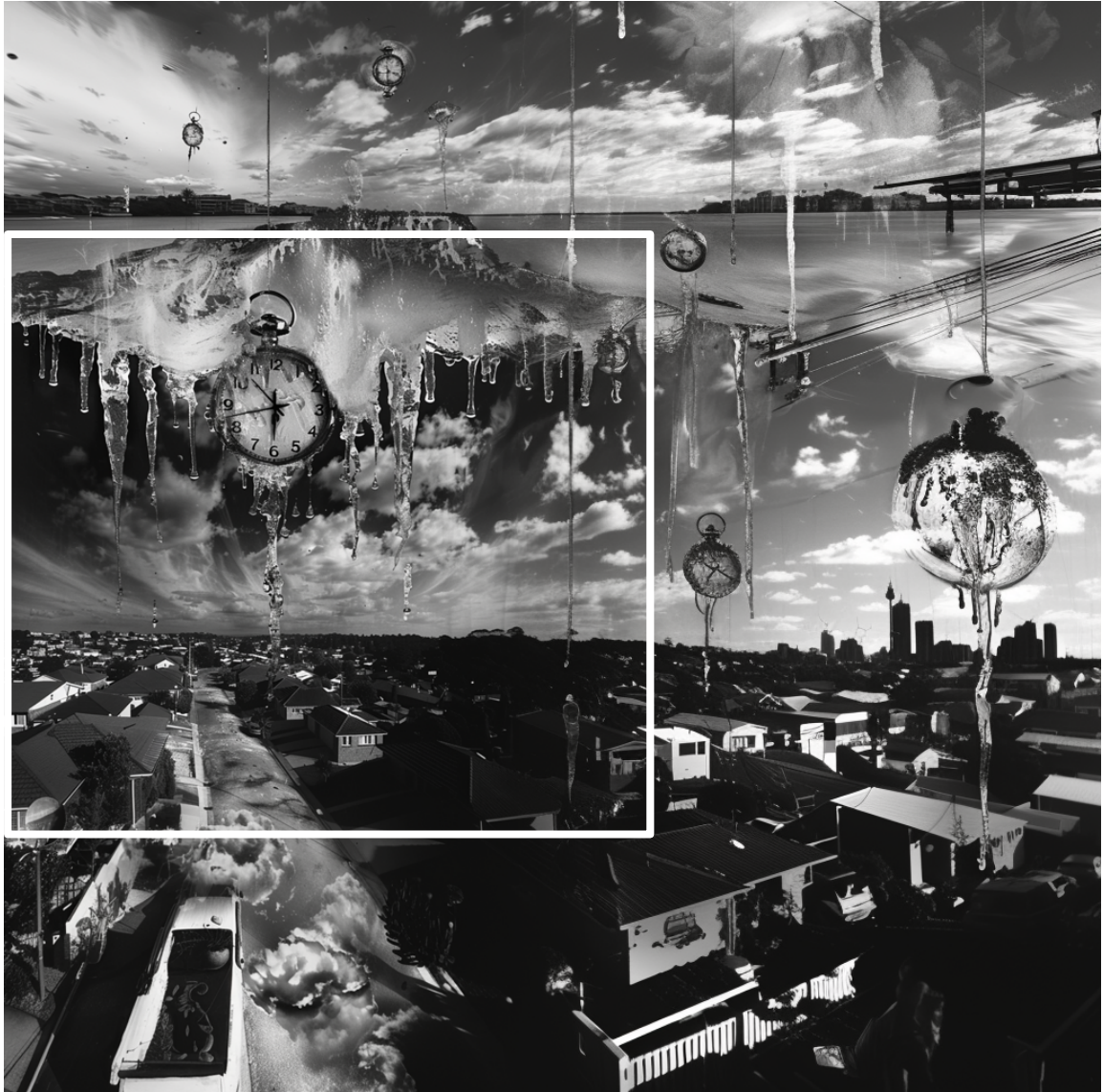

*Figure 4b*

Figure *4b* was a deliberate attempt at evoking a surreal image. In this case, rather than relying on the artist name to direct aesthetics and style, it is the phrase 'melting clock made of ice' that evokes an iconic form reminiscent of Salvatore Dali, superimposed onto a photo of a typical Australian suburbia. The surreal tendency in the inset, when prompted, using the pan command (in three directions: top, right, bottom) introduced most significantly, a second horizon line, coinciding with the top of the original image. Hanging from both the original and the new horizon lines are a series of melting clocks of different scales in the foreground and the background. The new horizon line creates an ambiguity in the image between sky and water, so that what originally was the sky dotted with clouds in the inset expands to the new image as a sky merging into water and then a shoreline. Further subjects are introduced, the suburb's street extends with textures resembling clouds on the ground, while in the background of the extended suburb is the skyline of a distant metropolis. In place of a proper name, the improbable but famous juxtaposition ('melting clocks') grants latitude to the model's associations, and links to other surreal effects. Unlike *4a*, which introduces the fantastical into an image with an underlying structure of a photograph, *4b* modifies composition as well as individual elements.

# From Prompt to Pixel: Attentive Media

Of the many moments of genesis of today's generative AI, one of the more significant is the 2017 publication of *Attention is All You Need*, a paper by a group of Google engineers proposing a new architecture for machine learning (Vaswani et al., 2017). Aside from belonging to a genre of computer science articles with vogue-ish subject-predicate titles, one of paper's technical contributions is the idea of learning the connections between a word (or 'token' – something like morphemes) to every other word in a sentence or phrase. 'Attention' is this act of association, creating networks out of lists of words, and assigning weights to each link of the network that conditions how often that link occurs. The surprise result of this 2017 paper was that the attention mechanism is 'all you need' to produce highly coherent text completion, paving the way the generalised pre-trained series of attention-based models that are foundational for generative AI. Our experiments with constrained syntax shows how this attention can be read as a form of intensive inward looking – an in-visuality, as Mackenzie and Munster's term suggests, or more generally, an in-mediality.

It seems an anachronistic jump to translate talk of weights and tokens to the now-quaint world of Saussurian signifiers, but in doing so we notice these language models are adept in the key signifying operations of *selection* and *combination*. In Lacan's further transpositions, selection corresponds to the rhetorical figure of metaphor and the psychoanalytic act of condensation, while combination corresponds to the equivalent pair of metonymy and displacement. In the 'Seminar on "The Purloined Letter"' (2001), he had drawn explicit connections between these acts and those involved in the construction of Markov chains: sequences of signifiers follow probabilistic rules that govern the text term in the sequence. These simplified Markov chains – critical precursors to today's generative AI – hold for Lacan a 'recreational character' that nevertheless can 'allow us to conceptualise where the indestructible persistence of unconscious desire is situated' (Lacan, 2001, p. 52). The rules as well as the networks of signifier that govern productions of speech remain in the background, in the latent space that forms an individual subject's unconscious, and it is desire, and the directing of attention needed to satisfy it, that mobilises each articulation, each process of selection and combination according to a background symbolic order into which the subject is drawn.

One of the difficulties then of criticising the unthinking operations of generative media is that at least according to some descriptions of communication, this is already what human productions of speech already do. The accelerated comprehension of today's systems remains, certainly, a trick of a 'stochastic parrot' (Bender et al., 2021), but it is one that with each new AI iteration seems less removed from the tricks that we as originary users of language already employ. The iterated application of guidance that has steered generative language models towards greater public acceptability only reinforces the impression that these models imitate some degree of social norm compliance alongside the rules of syntactic composition, that they pay attention not only to the immediate words present in a given context, but to the multiple registers of language – social, pragmatic, semantic, syntactic and morphological – in parallel (Magee, Arora & Munn 2023). This impression is in turn misleading; there is no 'outside' of the text, image or model, no actual world or society of others that separately anchor meaning. But even if the mechanism differs, it remains comparable to the kinds of 'repetition automation' that psychoanalysis identifies in sites – dreams, jokes, parapraxes – of unconscious language acts.

What can be more easily distinguished is rather the foundational desire that initiates generative power in both human and machine – if we understand by 'machine' here an advanced form of what Stiegler (2019) termed 'tertiary retention', the exteriorisation of memory. Even the apparently agentive character of generative AI is a distillation, compression and refraction of vast reservoirs of human desire, rippling across everything from accumulated databases of culture to the concentrations of capital that fund data centres, engineer salaries and the thousands of graphic cards needed to instil the digital contours that guide these outputs. This desire can still at this time only be spoken about in relation to the machine metaphorically; it remains at its core a human desire, unconscious and contradictory, that directs the machine's attention, right down to the final consumer's design of a prompt. It is as though the generative model had first fractured a collective cultural human memory – distorted, always, by who captures and who is captured in this memory, at what proportions, and so on – then, on demand, reassembles these fragments into some new memory that no one had. What it is reconstituted is an attention or even obsession towards a non-object that still, with its unrealness, can disturb, confront, or haunt the observer.

In moving from speech to vision, from the mouth to the eye, we obtain another perspective about how this attention spirals out from a set of minute technical operations to a constitutive gaze that, inhuman, nevertheless as the synthesis of nearly all visual memory also looks out or back upon the human subject, paying it all the attention it wants, at least in the literal sense of responding to that subject's request. How does this interpolation work? We can describe first how the generative image gets our attention, teasing out moments of 'surprise', constitutive of the gaze. Lacan writes (1977, p. 84):

> The gaze, as conceived by Sartre, is the gaze by which I am surprised—surprised in so far as it changes all the perspectives, the lines of force, of my world, orders it, from the point of view of nothingness where I am, in a sort of radiated reticulation of the organisms.

In an Anthropocenic world saturated by the visual, it could already be asked whether this formulation, after the advent of cinema and photography but prior to the ubiquity of television and the digital screen, already seems quaint. Can we be surprised any longer, in anything like this dramatic sense? If not in the immediate sense of apprehension, we still can imagine a more subdued surprise arising from the frequent distortions effected by the prompt.

In the first set we had not anticipated the name of a photographer alone being sufficient to condition, more or less, the historical period of Sydney's urbanisation mapped through its shifting skyline accompanied by a continual evolution of urban morphology and style. In the second set – and admittedly these had been selected on the basis of an immediate striking contrast – we can highlight the extent to which a simple lexicon can retrieve complex contrastive arrangements of subject and style. The drones swarm over the beach ominously (2a), while the (feminised) eye regards a similar scene (2c) with – what? some mix of wonder and apprehension? The clause 'dignified poses', picked up, alongside others, from examples from the Midjourney community Discord server, seems more likely to introduce human figures into the image, and to add to the sombre qualities of the black and white palette.

In Lacan's treatment of the gaze (2007), the viewer is interpolated by what is viewed. 'The picture, certainly, is in my eye. But I too am in the picture' (Lacan, 1977, p. 96 – own translation): thrown into the picture or tableau, the seeing subject recognises itself as an object that appears in the eyes of others, caught in the act of framing. We thought it would be interesting to experiment with this expression itself, to capture an image of a machine staring back – partly a test of how the machine represents theories of reflexivity. This proved difficult: image *2b* for example insists upon tracing an imprint of a white male version of 'spectral humanoid bot', either posing or caught unaware, an accident of photographic technique. In our iterations we could not coerce Midjourney into an explicit *machinic* self-portrait; our sampling would either reproduce the white male photographer staring back, or a robot in ¾ or full profile (*2d*). This suggests – we cannot know for sure – a strange and latent semantic axis of 'androgynous-robot-portrait' / 'white-male-human-self-portrait' fissuring these representations. The machine can reproduce a non-machine seeing itself and it can reproduce a machine being seen, but it cannot *see itself in a mirror*.

The third series of images ruminate more directly on the introjection of the Anthropocene into an Australian imaginary. Image *3c* shows how signifiers (diver and waste) can collapse into a metaphorical relation – the waste *is* the diver – even though otherwise little is left in the prompt to coerce a surrealistic leap. In the second set (*3e-h*), this metaphorisation is more pronounced. The bottle is no longer waste but life-support (*3e, 3h*); or diver and bottle become both cause and part of the decaying reef (*3f, 3g*). The intricate play between word and image is extended in the fourth series, where pan and zoom operations include the original image as part of the prompt. Image *4a* illustrates how these operations do not simply extrapolate, but also interpret and speculate. The flying sheep are nowhere in the original, but both flying objects and sheep are, and again they are condensed into multiple fantastic objects. The figure of the observing 'farmer' is also replicated, as though the act of zooming out offers opportunity for a kind of meta-commentary on the original frame, a machinic reinterpretation of the infinite mirror image. Image *4b*, finally, is the most direct example of an anamorphic effect, as the zooming out produces a duplicate wide–angle landscape at the top of the image. What appears as an artefactual error still blends in just plausibly enough, as the sea becomes sky and street becomes water.

But these images do more than surprise. They are also active; even in the strangest cases there is something seen in the history of art and photography that is speaking back. In her discussion of images of the Arctic, video artist Susan Schuppli talks of how technical operations *intensify*:

> I love that moment because you can take something that is very expansive and even sometimes ordinary looking and then, by framing that or shifting focus or using a different lens, you kind of reinvent that space or produce an intensification of that larger image (Lee-Morrison, 2023).

Generative reproductions act as related but different technical instruments of intensification. Decomposed into the model's set of features, during the generative moment these other views cohere into a new social ensemble – a vocalisation or visualisation that in some non-conscious way articulates a history of past and now dead attention. The prompt becomes an incantation that calls upon this visual archive to state: this is what humans have been paying attention to. Broken down into tokens or clauses, the text recalls pictorial fragments extracted from an expanding archive of memories, visions, apprehensions. Salvaggio (2024) describes how these image generators are 'surveillant', part of a capitalist enterprise that appropriates the history of the image in order, within an expanding and competitive attention economy, to commoditise and intensify it. Serving the consumer involves seizing their attention; the obsequious bot underhandedly asserts its mastery. Consumer feedback is captured as gestural nudges that improve future models, and in this sense the consumer is also 'in the picture'.

But in drawing upon a history of attention, the composition also places the viewer in a relation to this history, and demands this relation also be paid attention. There is in other words a scopic symmetry at play when the same viewer writes a prompt and peers over the result – this scene has itself already been internalised by the machine, which is nothing more than a decomposition of the same subject-object relation, over and over. The act of paying attention is itself traced in the coordinates of the generated image. This recursive operation of an observer caught in what is observed belongs to a wider Anthropogenic condition in which the human constitutes their environment. The manipulations wrought by the machine accentuate, and act as oblique commentary upon, the productive qualities of the gaze.

# Conclusion

We suggested earlier the figure of the latent constellation, to describe how the clauses of the prompt act separately or in concert to draw out associated pixel groups. The importance of this trope is in seeking a position that neither accedes entirely to the critiques of averageness, nor buys into the hype of some newfound technical agency. Rather we want to think of a prompt as a structured discursive act, one that cuts into what is a redistribution – or a 'radiated reticulation' (Lacan) – of signification and desire, accumulated in the digitalised archiving of text and image. The gaze of this machine is retrospective, or even retro*jected*, thrown back into a digital equivalent of a dusty windowless archive, a dimly lit rather than entirely black box, with only facsimiles to shuffle through and a never-ending stream of requests from some kind of voiced exterior demand. The Anthropocene, to which with its vast consumption of rare earth minerals and emissions it only too eagerly contributes, is at the same time a myth, only one motif that binds elements in a vast repertoire of signifying elements. If for Lacan, normal or neurotic human subjectivity depended upon the coordination of a set of master signifiers which fixate, and accordingly coordinate, our daily attention spans, the generative model is distinguished by the easy reprogramming of its attention – precisely in fact towards those same obsessions that direct *human* attention.

Yet it does so through a lens that precisely in its distortions can collide affective dimensions of human desire. The tragic or sublime is mixed in with the whimsical and farcical. The photographic 'witness' becomes compromised with distracting aesthetic effect – introduced, to be sure, as part of our explorations. What ought to be simple technical operations of panning and zooming inadvertently shift genre, from realism to surrealism. Tricks, such as the 'double exposure… superimposed on' bracket, expose both fluencies and rigidities in the archive. And so on. In conjunction these permit less a radically alternative than oblique reading of visual representations of the Anthropocene, or generally, of what from at least the middle of the twentieth century was an acknowledged destructive and schismatic relationship between human and world.

What are we to say in relation to this gaze – like and yet utterly unlike the history of technical media that precedes and conditions it? Is it sufficient for example to label these systems as 'generative AI' or 'synthetic media', which seem to do little justice to quite specific ways they organise word tokens or image pixels into coherent ensembles? Or more radically: do the operations of these systems replicate *in silica* key artistic operations of selection, arrangement and distinction that would otherwise remain the preserve of human perception and culture? And if distinct at the level of *genus* – if, as Parisi (2019) suggests, the machine's view is 'alien' – does there remain in this distinction something that points productively back upon interpretations of human culture? In exploring combinations of a deliberately controlled set of variations, our argument has less been concerned with circumscribing the boundaries of representations of something like the Anthropocene or Australian visual culture. Instead it has been to highlight a novel phase of a general media environment that is commensurate with the Anthropocene, politically as much as chronologically. If this machinic gaze extracts and surveils, we also want to acknowledge a potential for experimentation and engagement with a history of the image that, unlike the linearity of keyword searches or index cards, is stitched out of the strange combinatorial possibilities of the prompt.